\documentclass{optica-article}

\journal{opticajournal} 

\articletype{Research Article}
\usepackage{graphicx}
\usepackage{soul}

\begin{document}

\title{Spectral Speckle Customization}

\author{Nicholas Bender,\authormark{1,*} Henry Haig,\authormark{1} Demetrios N. Christodoulides, \authormark{2} and Frank Wise\authormark{1}}

\address{\authormark{1}School of Applied and Engineering Physics, Cornell University, Ithaca, New York 14853, USA\\
\authormark{2}Ming Hsieh Department of Electrical and Computer Engineering, University of Southern California, Los Angeles, California 90089, USA\\
}

\email{\authormark{*}nicholas.bender@cornell.edu} 


\begin{abstract*} 
Speckle patterns are used in a broad range of applications including microscopy, imaging, and light-matter interactions. Tailoring speckles’ statistics can dramatically enhance their performance in applications. We present an experimental technique for customizing the spatio-spectral speckled intensity statistics of optical pulses at the output of a complex medium (a disordered multimode fiber) by controlling the spatial profile of the input light. We demonstrate that it is possible to create ensembles of independent speckle patterns with arbitrary statistics at a single wavelength, simultaneously at decorrelated wavelengths, and even tailored statistics across an entire pulse spectrum.
\end{abstract*}

\section*{Introduction}

Speckle formation is ubiquitous, irrespective of wave-medium or even wave-nature. Whenever a large number of random partial waves interfere, the result is an irregular granular pattern known as a speckle pattern. If the random partial waves are statistically independent, with uniformly distributed phases between 0 and $2\pi$, the resulting speckle pattern is called “fully-developed”. While the likelihood of finding two identical fully-developed speckle patterns in nature is effectively zero, nearly all speckle patterns obey Rayleigh statistics. Typical Rayleigh speckles have a circular Gaussian field probability density function (PDF), a negative exponential intensity PDF, $P(I/\langle I \rangle) = \textrm{exp}[-I/\langle I \rangle]/ \langle I \rangle$, and lack any significant spatial coherence outside the diffraction limited speckle grains~\cite{DaintyBook, GoodmanBook, freund1001}. Thanks to both their near-universal statistics and their pervasiveness, speckles have been utilized in numerous practical applications. In optics, for example, speckle patterns have been used in imaging/wide-field microscopy applications~\cite{ventalon2005quasi, ventalon2006dynamic, shapiro2008computational, lim2008wide, bromberg2009ghost, katz2009compressive, lim2011optically, bernet2011lensless, mudry2012structured, min2013fluorescent, yilmaz2015speckle, phillips2016non,chaigne2016super, singh2017exploiting, vigoren2018optical, yeh2019speckle, pascucci2019compressive, zhang2019near, choi2022wide, affannoukoue2022super,zhu2022large}, as well as in speckle-based sensing techniques~\cite{fercher1981flow, pan1994multimode, briers1996laser, briers2007laser, kim2016remote, berto2017, murray2019speckle, baek2020speckle, liu2020spectral, luo2021super}. Speckles have been adapted for use in optical trapping and micro-manipulation~\cite{shvedov2010selective, jendrzejewski2012three, volpe2014speckle, kotnala2020opto}, as well as to control living cells in active media~\cite{douglass2012superdiffusion, bechinger2016active, pincce2016disorder}.

Even though Rayleigh speckles are the nearly universal family of speckles found in nature, over the years, attempts have been made to create non-Rayleigh speckles~\cite{fujii1974effect, goodman1975dependence, fujii1975statistical, jakeman1984Gaussian, levine1983non, Jakeman1984Speckle, bromberg2014generating}. Early research demonstrated that non-Rayleigh speckles arise when the speckled field is either under-developed (not fully randomized or the sum of a small number of partial waves) or partially-coherent (the sum of incoherent partial waves)~\cite{fujii1974effect, goodman1975dependence, fujii1975statistical, jakeman1984Gaussian, levine1983non, Jakeman1984Speckle}. After the advent of liquid crystal based wavefront shaping, a straightforward technique for creating fully-developed non-Rayleigh speckle patterns in free space was developed~\cite{bromberg2014generating}, where the non-Rayleigh speckles could either adhere to an intensity PDF with a tail decaying faster (sub-Rayleigh) or slower (super-Rayleigh) than a negative-exponential function. Subsequent to this, a method for creating fully-developed speckle patterns with \textit{arbitrary} intensity PDFs was developed~\cite{bender2018customizing}. These, and other contemporary works~\cite{bromberg2014generating, dogariu2015electromagnetic, fischer2015light, amaral2015tailoring, kondakci2016sub, di2016tailoring, guillon2017vortex, bender2018customizing, di2018hyperuniformity, bender2019introducing, bender2019creating, bender2021circumventing, devaud2021speckle, liu2021generation, han2023tailoring}, opened the door to many practical applications involving bespoke speckle patterns. Introducing tailored speckles into standard speckle imaging applications can improve the signal-to-noise, resolution, and/or the overall image quality~\cite{oh2013sub, zhang2016high, li2021sub, liu2019label, pascucci2019compressive, zhang2015ghost, kuplicki2016high, nie2021noise, GuilbertPHD}. For example, in the context of parallelized nonlinear pattern-illumination microscopy (STED, RESOLF, etc.) tailored speckles can produce a spatial resolution three-times higher than the optical diffraction limit, before an equivalent Rayleigh speckle approaches can become diffraction limited~\cite{bender2021circumventing}.

Generally, existing speckle customization techniques are intended for use in free space, with monochromatic or narrow-band light (nanosecond pulses), i.e. light lacking spatio-spectral complexity. On the other hand, broadband spatio-spectrally complex light is used in a number of applications~\cite{cao2022harnessing, mosk2012controlling}. For example, in snapshot hyperspectral imaging~\cite{french2017speckle, sahoo2017single, monakhova2020spectral}, the spatial or spectral scanning required in conventional multispectral imaging is replaced by a single image acquisition. This is accomplished by leveraging speckled spatio-spectral complexity to perform measurements in parallel. Bespoke speckles have the potential to enhance broadband speckle imaging applications like this, as they have done for monochromatic imaging applications. There is also a general interest in controlling short pulse behavior in complex media -- such as multimode fiber (MM) -- due to the potential applications in optical telecommunications~\cite{richardson2013space}, and MM high-powered pulsed lasers~\cite{wright2015controllable, wright2017spatiotemporal, haig2022multimode}. Current speckle customization techniques cannot be applied to these research areas, however, because they typically involve broadband pulses which can be spatio-spectrally complex. Furthermore, it is not known if the speckle statistics in a broadband pulse can be controlled, and if this is possible, to what extent. Even in the context of monochromatic speckle customization, important questions remain unanswered. Whether it is possible to arbitrarily customize the intensity statistics of speckle patterns after a complex medium remains unknown. Along these lines, the influence of common effects like incomplete control --due to either imperfect or insufficient light coupling into a medium, as well as index mismatch at the surfaces-- on controlling speckles at the output of a complex medium remain unknown.

In this work, we experimentally demonstrate the ability to customize the intensity statistics of spatio-spectrally speckled pulses at the output of a disordered MM fiber. To accomplish this, we leverage the ability of a disordered MM fiber to convert spatial degrees of freedom -- controlled using a phase only spatial light modulator modifying the light entering the fiber -- into spatio-spectral degrees of freedom, which are in turn used to customize the speckle statistics across the pulse. Because the spectral decorrelation length of the fiber ($4$ nm) is smaller than the spectral width of the pulse ($20$ nm), we can generate independent speckle patterns with distinct statistics at different wavelengths within the pulse, using a single SLM pattern. We show that it is even possible to create non-Rayleigh speckles across the entire spectrum of the pulse. To create these distributions, we start with an ensemble of speckled pulses output from the fiber and tailor the spectrally-dependent statistics of the speckles using sequential PDF transformations regularized by a spatio-spectral field-mapping matrix that characterizes the MM fiber.

\section*{Experimental Setup}

\begin{figure}[htb]
    \centering
    \includegraphics[width=\linewidth]{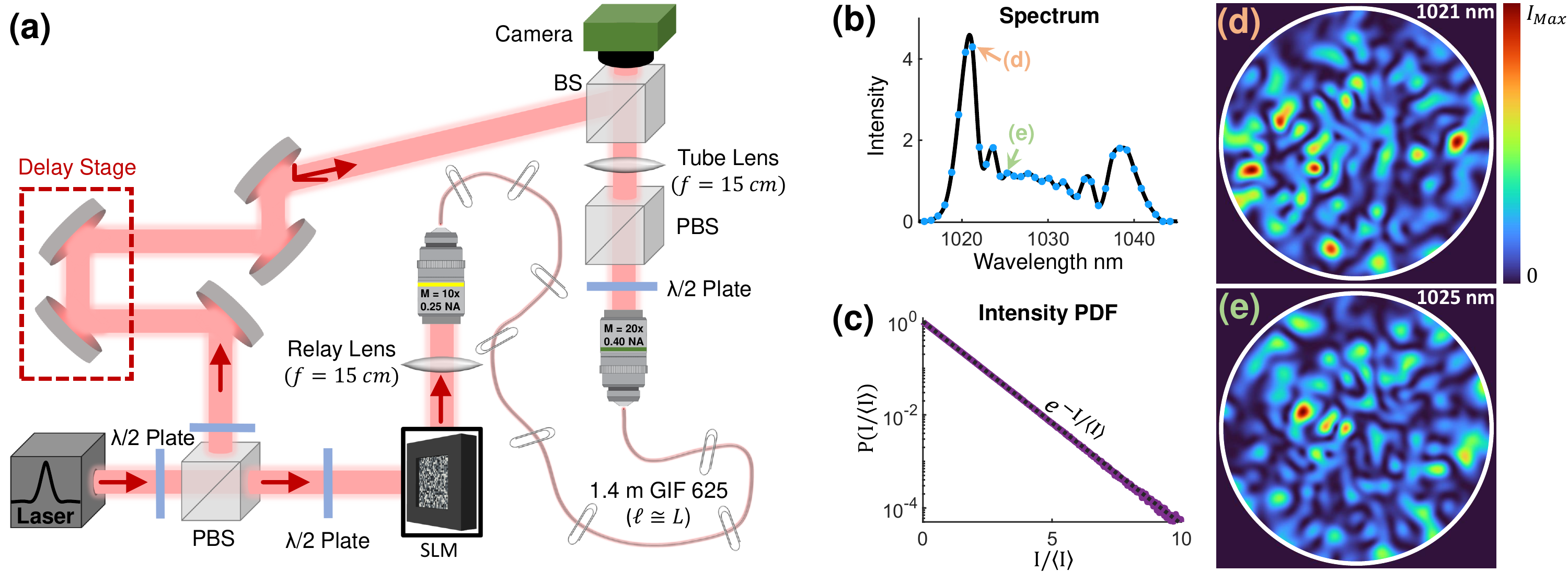}
    \caption{\textbf{Experimental setup (a) and standard pulse properties (b-e).} Light from a transform-limited pulsed laser source is split into two beams. One beam is wavefront shaped by a SLM and injected into a disordered MM fiber, while the other is used as an interferometric reference beam. The output of the disordered fiber is imaged by a camera, where the two beams are recombined and used to measure the pulsed field output from the fiber. A typical spatially-averaged spectrum, measured using the interferometer, is shown in (b). Under random input conditions, the normalized speckled intensity patterns output from the fiber (d,e) adhere to Rayleigh intensity statistics (c). The measure intensity PDF (purple line), follows the theoretical prediction for Rayleigh speckles (black dashed line). The spectral resolution (0.8 nm) is significantly smaller than the spectral decorrelation length ($4$ nm) of the pulse output from the fiber. In (d,e) the white circle marks the fiber core.}
    \label{Setup}
\end{figure}

The experimental setup used in this work is illustrated in Fig.~\ref{Setup}(a). An all-normal-dispersion fiber laser~\cite{chong2006all} and grating compressor  generate a train of approximately transform-limited pulses (temporal duration 150 fs full-width at half-maximum (FWHM), repetition rate 35 MHz, and representative average spectrum shown in Fig.~\ref{Setup}(b)). The pulses are split into two separate beams using a $\lambda/2$ plate and a polarizing beam splitter (PBS). One beam illuminates the surface of a phase-only spatial light modulator (Meadowlark P1920-1064). The liquid crystal display on the spatial light modulator (SLM) is fully illuminated and a circular array of 156 macro-pixels controls the light reflected from the panel (the circle radius is $\approx 7$ macro-pixels, and each macro-pixel consists of $64 \times 64$ SLM pixels). Within each macro-pixel a stepped diffraction grating is displayed to control the amplitude and phase of the light in the 1st-order diffraction of each macro-pixel. Using a $f = 15$ cm relay lens, we image the SLM macro-pixel array onto the back focal plane of an objective lens ($10\times$, NA = 0.25) -- via the 1st-order diffraction pattern -- which then excites a graded index multimode fiber (Thorlabs GIF 625) with the Fourier transform. The 1.4 m long fiber supports $\approx 325$ modes per polarization, and it is clamped at numerous locations along the fiber to introduce strong mode mixing. Specifically, the characteristic disorder length $\ell$ (often called the transport mean free path~\cite{xiong2017principal} or the equilibrium modal distribution length~\cite{garito1998effects}) is approximately the length of the fiber. As a result, effectively any input spatial distribution is completely randomized when it reaches the end of the fiber. We image a single polarization of light output from the fiber onto a CMOS camera (Allied Vision Alvium U-511) using an objective lens ($20\times$, NA = 0.40) with a $f = 15$ cm tube lens. The second beam is path-length-matched to the light imaged by the camera, and recombined with the light output from the fiber using a beam splitter before the camera. The combination of a variable delay stage along the beam path with the near plane-wave spatial profile on the camera enables off-axis digital holography to reconstruct the spatio-spectral field output from the fiber~\cite{Cuche2000Spatial}. With this setup, we measure the spatio-spectral field-mapping tensor $(\hat{x}, \hat{y}, \hat{\lambda})$, which can predict the full spatio-spectral field output from the disordered fiber for any input field modulation pattern created by the SLM macro-pixel array (see supplementary section 1). With this operator, we study the properties of light traversing the disordered fiber. 

\section*{Rayleigh Pulse Properties}

The spatio-spectral properties of light output from the disordered fiber under Rayleigh speckle illumination conditions (created by displaying uniformly distributed random phase patterns on the SLM array) are shown in Fig.~\ref{Setup}(b-e). The spatially-averaged spectrum of the pulse output from the fiber (Fig.~\ref{Setup}(b)) is unchanged relative to that of the input pulse. The intensity PDF of the normalized spectrally-dependent spatially-speckled fields is obtained after ensemble averaging 1000 independent realizations (purple line in Fig.~\ref{Setup}(c)). This PDF adheres to Rayleigh statistics (black line). To normalize the output intensity patterns we use $\overline{I(x,y)/\langle I(x,y) \rangle}$ = 1, where $\langle~\cdots~\rangle $ denotes ensemble averaging and the overline indicates spatial averaging. This normalization is always used unless stated otherwise. For analysis of the statistics without normalization and effects on the method presented in this work, see supplementary section 2. Visual inspection of the output spectrally-dependent spatial-intensity patterns (Figs.~\ref{Setup}(d,e)) indicates that the speckle patterns generated at the output of the disordered fiber are fully-developed. This observation is rigorously confirmed in supplementary materials section 3.

One of the important properties of the speckled light output from the disordered-fiber, relative to what can be created with a SLM and free-space optics, is the spatio-spectral complexity of the pulse. This property can be quantified by comparing the FWHM of the magnitude-squared spectral field-correlation function $|C_E(\Delta \lambda)|^{2}$ to the pulse bandwidth ($20$ nm). The spectral correlation function is defined as
\begin{equation} 
C_{E}(\Delta \lambda) \equiv \frac{\langle E({\bf r}, \lambda) E^{*}({\bf r}, \lambda + \Delta \lambda)\rangle}{\sqrt{\langle |E({\bf r}, \lambda)|^2\rangle}\sqrt{\langle |E({\bf r}, \lambda + \Delta \lambda)|^2\rangle}},
\end{equation}
where $\langle ... \rangle$ denotes both spatial/spectral averaging over $(x,y,\lambda)$ and ensemble averaging over different speckle realizations. For the case of a spectrally-homogeneous speckle pattern, such as those used to excite the disordered fiber, the spectral correlation width (denoted $\textrm{FWHM}[|C_E(\Delta \lambda)|^{2}]$), greatly exceeds the spectral bandwidth; i.e., each wavelength in the pulse has approximately the same speckle pattern. In contrast, with the disordered fiber in the system the spectral correlation width, $\textrm{FWHM}[|C_E(\Delta \lambda)|^{2}] = 4 \textrm{ nm}$, is significantly smaller than the pulse bandwidth. As a result, for a single spatial input pattern, the output pulse spatial pattern completely changes for spectral components with a separation of 4 nm, as shown in Figs.~\ref{Setup} (b,d,e). In other words, the pulses output from the disordered fiber are $5\times$ more spatio-spectrally complex than those at the input of the fiber, assuming an equivalent number of modes are excited. Since the stated goal of this work is \textit{controlling} the output pulse properties, it may seem counterintuitive to increase the complexity of the output pulse -- relative to our ability to control the input -- by introducing the disordered fiber. This has the advantage, however, of enabling us to convert purely spatial control over the pulse properties into spatio-spectral control over the pulse properties, at the expense of the overall ability to control the pulse.

\section*{Method}

Our goal is to customize the intensity PDFs of the spectrally dependent speckles output from the fiber, and the technique for accomplishing this is an adaptive local-intensity transformation algorithm based on spatio-spectral field mapping.  

\subsection*{Field Mapping Matrix}

The first step in this approach is to define an appropriate field-mapping matrix that relates the macro-pixel array on the SLM to the spatio-spectral fields we wish to control at the fiber output, and to construct a matrix representing the inverse relationship. While the array on the SLM is two-dimensional, we represent it as a 1D vector by assigning each macro-pixel’s field modulation to an index, $n$, out of $N$ total ($N=156$ for our system). Similarly, we represent the output at each spectrally-dependent camera pixel of interest $(x, y, \lambda)$ using a single index, $m$, out of $M$ total. Here $M$ is the number of camera pixels measuring the light in the core of the fiber, $M_{core}$, multiplied by the number of different spectrally-dependent fields we are attempting to control, $M_\lambda$. It is worth noting that the unperturbed fiber mode basis may seem more appropriate than the spatial basis. In practice, there are two main drawbacks to using the fiber mode basis. First, with only 156 macro-pixels available on the SLM, we don’t have complete control over the $\sim325$ single-polarization modes at the input of the fiber. Second, to calculate the intensity PDF we need the real-space representation of the field. On the other hand, there is one restriction on using a spatially-indexed representation of the fields: the spatial sampling grid ($x,y$) must satisfy the Nyquist sampling theorem. Experimentally, we sample the fields at the output of the fiber at a rate of $\sim18$ points (along a single dimension) per average speckle grain width. We can therefore represent the field-mapping from the SLM macro-pixels to the camera pixels using a rectangular ($M \times N$) matrix using this basis and experimentally measure the field-mapping tensor to determine the matrix elements. Because it is rectangular, the field-mapping matrix is not guaranteed to have an inverse. Nevertheless, we can still calculate its Moore-Penrose inverse matrix~\cite{penrose1955generalized}, often called the ``pseudo inverse” matrix. To a good approximation, applying the pseudo-inverse matrix to any field created by applying a SLM pattern to the field-mapping matrix results in the SLM pattern. When the pseudo-inverse is applied to an arbitrary/artificial field, it gives a ``best guess” for the SLM pattern that generates the field. However, the obtained pattern will not necessarily generate the arbitrary field. Therein lies the need for our method to be adaptive.

\subsection*{Adaptive PDF Transformation}

Our next step is to define a formalism for obtaining a function which transforms a set of scalar intensity values, $I(x, y, \lambda)$, adhering to a given intensity PDF, $P(I)$, into a new set of scalar intensity values, $J(x, y, \lambda)$, adhering to the desired intensity PDF, $F(J)$. This type of local intensity transformation~\cite{ bender2018customizing, bender2019creating, han2023tailoring} is obtained by equating an infinitesimally narrow area element in each PDF, $P(\tilde{I}) d\tilde{I} = F(\tilde{J}) d\tilde{J}$, integrating
\begin{equation}
\label{int}
\int_{0}^{I} P(\tilde{I}) d\tilde{I} = \int_{0}^{J} F(\tilde{J}) d\tilde{J},
\end{equation}
and solving the integral equation to write $J$ as a function of $I$. Generally, it is a good practice to ensure that both PDFs have the same normalization, i.e., $\int_{0}^{\infty} P(I) dI =1$, and the same average intensity value $\langle I \rangle =1$. Equation~\ref{int} can be numerically solved for any physically allowed $F(J)$ and $P(I)$, and the solution $J(I)$ is the desired local intensity transformation. Certain intensity PDFs are unphysical, however. For example, a zero in the PDF ($P(I) = 0$, when $0<I<I_{max}$) would necessitate a discontinuity in the field and is impossible for a classical light field. One of the advantages of using a local intensity transformation to alter the intensity PDF of a speckle pattern, or an ensemble of speckle patterns is that it is a scalar function. As such, if the original ensemble of speckles consists of statistically stationary and ergodic random variables, which is the case for fully-developed Rayleigh speckles, then the transformed variables are also stationary and ergodic~\cite{bender2018customizing}. 

\subsection*{Iterative Process}

With the field-mapping operators defined and an adaptive PDF transformation technique identified, we can iteratively use the two to arbitrarily customize ensembles of speckles. We start by creating an ensemble of 100 independent speckle realizations, each generated by applying a different random SLM phase array to the field mapping matrix. The intensity PDF of the entire ensemble is then calculated. The obtained intensity PDF is used in conjunction with the desired PDF to obtain a local intensity transformation function. The local intensity transformation is applied to the amplitude values of the field in each speckled realization, transforming the ensemble’s intensity PDF to the desired PDF. The phase value associated with each amplitude value is left unchanged. The result is an ensemble of ``artificial" fields that are not necessarily possible to generate with the SLM, as discussed previously. Applying the pseudo-inverse matrix to each transformed field yields a best guess for a generating SLM field. By reapplying the field mapping matrix to each best guess we obtain an ensemble of non-artificial fields. The resulting fields don’t exhibit the desired statistics; however, they also don’t follow the original intensity PDF. In fact, they are incrementally closer to the desired statistics. By iterating this process of generating a large ensemble of speckle realizations using independent SLM patterns, applying a local intensity transformation, applying the pseudo-inverse, and repeating with the new SLM patterns, we eventually obtain a set of independent SLM patterns which generate an ensemble of speckle realizations adhering to a desired intensity PDF at the output of the fiber. This method can be modified to generate different PDFs (at different wavelengths, for example) by applying a different intensity transformation to different elements (wavelengths) of a given speckle realization.

\section*{Results}
\subsection*{Monochromatic Customization}
\begin{figure}[hthb]
    \centering
    \includegraphics[width=\linewidth]{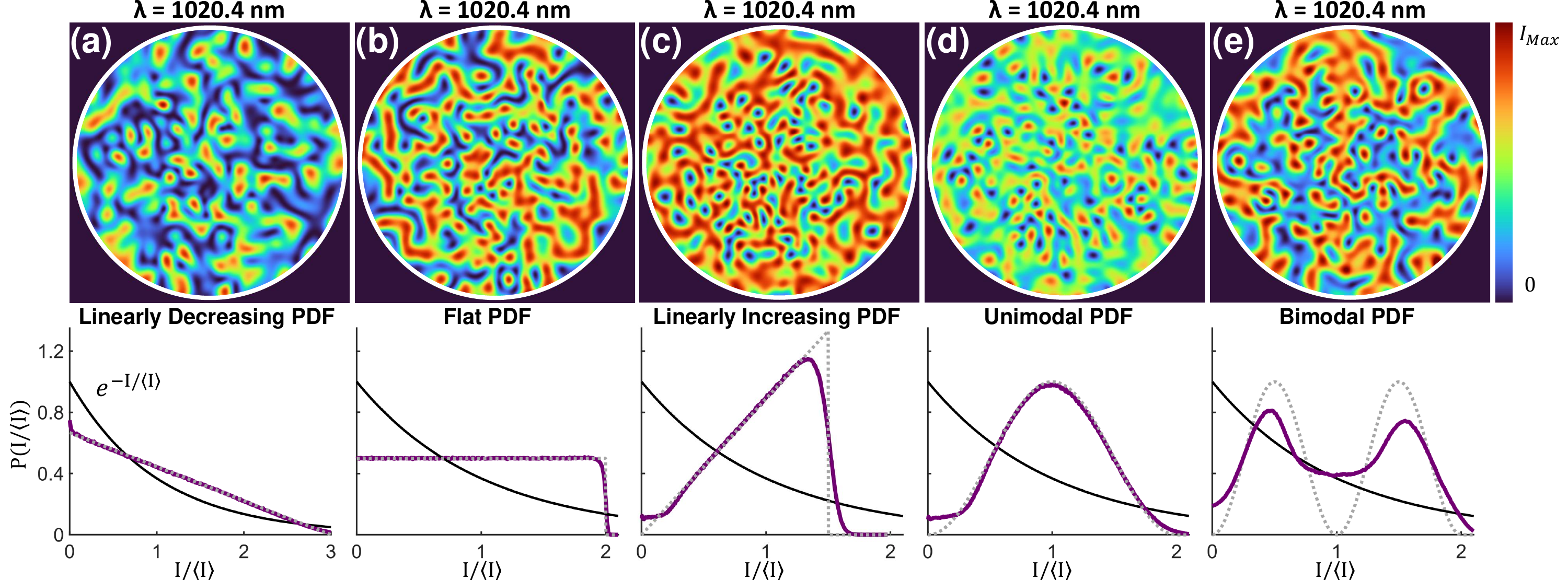}
    \caption{\textbf{Monochromatic speckle patterns with tailored intensity PDFs output from a disordered fiber.} Output speckle patterns adhering to linearly decreasing (a), uniform (b), linearly increasing (c), unimodal (d), and bimodal (e) intensity PDFs are shown. Below each output pattern, the associated ensemble-averaged intensity PDF (purple line) is compared with the ideal PDF (gray dots), and a Rayleigh PDF (black line).  The white circles are the boundary of the fiber core. One hundred independent speckle patterns are used to calculate each PDF, and the experimental data are obtained from the measured field mapping matrix.}
    \label{MonoPDF}
\end{figure}

Using the method described in the last section, we customize the monochromatic ($\lambda = 1020.4$ nm) intensity statistics across the entire fiber core. In Fig.~\ref{MonoPDF}(a) we present an example speckle pattern that adheres to a linearly decreasing intensity PDF, $P(I/\langle I \rangle)=(2/9)(3-I)$, over a finite intensity range, $0\leq I/\langle I \rangle \leq 3$. Unlike a Rayleigh PDF (black line) which monotonically decreases over an infinite range of intensity values, the custom intensity PDF (purple line) monotonically decreases over a finite range and then is zero for larger values of $I$. The monotonically decreasing nature of the of the custom PDF preserves the spatial structure/topology of a typical Rayleigh speckle pattern, an interconnected sea of low-intensity values surrounding isolated bright intensity islands, while the truncation of the custom PDF regularizes the peak intensity of the speckle grains. This can be seen by comparing Figs.~\ref{Setup}(d,e) with Fig.~\ref{MonoPDF}(a). Changing the intensity PDF from decreasing to uniform, $P(I/\langle I \rangle)=(1/2)$, over finite intensity range, $0\leq I/\langle I \rangle \leq 2$, changes the speckle topology but still regularizes the high intensity grain values, as shown in Fig.~\ref{MonoPDF}(b). For this custom PDF, the speckle grains irregularly merge, breaking apart the interconnected low intensity channel structure. By customizing the intensity PDF to linearly increase, $P(I/\langle I \rangle)=(8/9)(I)$, over a finite intensity range, $0\leq I/\langle I \rangle \leq 1.5$, the spatial-structure is inverted relative to Rayleigh speckles, as shown in Fig.~\ref{MonoPDF}(c). Here the speckle pattern consists of a sea of interconnected high-intensity grains surrounding isolated optical vortices. The interconnected structure in the speckle pattern need not be at the minima or maximal intensity values, as shown in Fig.~\ref{MonoPDF}(d). Here the intensity PDF is unimodal, $P(I/\langle I \rangle)=sin(\pi I/2)^{2}$, over finite intensity range, $0\leq I/\langle I \rangle \leq 2$. As a consequence, the average intensity values form a conjoining web while the vortices and the high-intensity regions remain isolated. Introducing a second maximum, as in the bimodal intensity PDF case shown in Fig.~\ref{MonoPDF}(e), creates a second intensity web structure. It is worth pointing out that Fig.~\ref{MonoPDF}(e) is an example of what happens when an unphysical intensity PDF, $P(I/\langle I \rangle)=sin(\pi I)^{2}$ over $0\leq I/\langle I \rangle \leq 2$, is used in our algorithm. The impossible zero in the intensity PDF is removed, and the exact PDF is not obtained when the iterative algorithm converges. The speckle patterns shown in Figs.~\ref{MonoPDF}(a-e) are representative examples taken from ensembles of 100 statistically-independent speckle patterns, all of which adhere to the purple intensity PDFs shown in the bottom row of Fig.~\ref{MonoPDF}. The speckles in each ensemble are fully-developed, stationary, and ergodic (see supplementary section 3). Taken together, Figs.~\ref{MonoPDF}(a-e) indicate that the intensity PDF of the speckles can be arbitrarily tailored. While this type of arbitrary customization has been shown before in free-space~\cite{bender2018customizing} this is the first demonstration after a disordered medium, and it has been accomplished without dividing the output into ``target” and ``junkyard” regions; the full output intensity pattern is customized. Finally, while the customization shown was performed on spatially normalized intensity values $I(x,y)/\langle I(x,y) \rangle$, in supplementary section 2 we show that the same results can be obtained without spatially normalizing the output.

\subsection*{Spectral Detuning}

\begin{figure}[hthb]
    \centering
    \includegraphics[width=\linewidth]{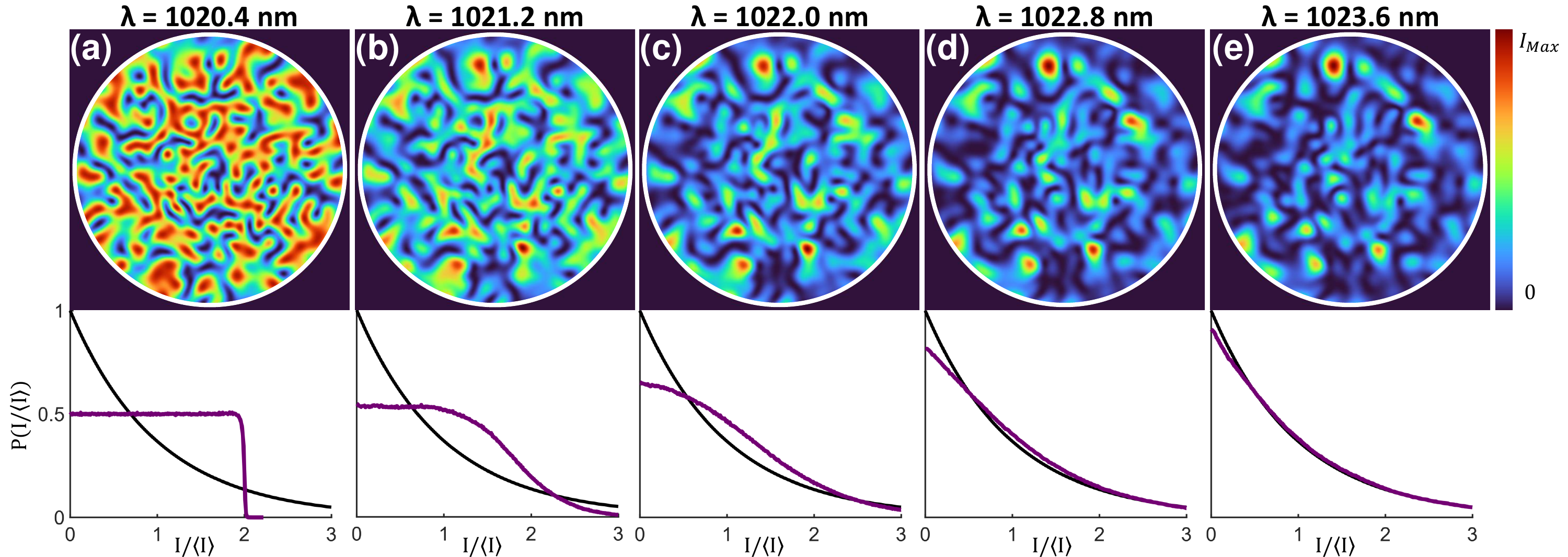}
    \caption{\textbf{Spectral detuning of customized speckles.} The output from the fiber, associated with a customized speckle pattern at $\lambda=1020.4$ nm, is compared to speckle patterns at spectrally-detuned wavelengths (b-e) generated by the same SLM pattern. Below each wavelength-dependent pattern, the associated ensemble-averaged intensity PDF (purple line) illustrates the speckles’ reversion back to Rayleigh statistics (black line) -- effectively obtained after (e) -- within one spectral decorrelation length, $4$ nm.  The white circles are the boundary of the fiber core. One hundred independent speckle patterns are used to calculate each PDF, and the experimental data are obtained from the measured field mapping matrix.}
    \label{Propagation}
\end{figure}

The tailored speckle PDFs of Fig.~\ref{MonoPDF} are obtained at a single wavelength. The pulse at the output of the fiber has a relatively broad bandwidth ($20$ nm) when compared to the spectral decorrelation length ($4$ nm), so the customized statistics do not persist across the spectrum. In Fig.~\ref{Propagation}, the spectral dependence of a customized speckle pattern is presented. Specifically, the case of a customized speckle pattern with a uniform intensity PDF (purple line) is shown in Fig.~\ref{Propagation}(a). For a small wavelength detuning of 0.8 nm (b), many of the speckle grains retain the same topological shape as in Fig.~\ref{Propagation}(a), but the speckle grain maxima are more irregular in Fig.~\ref{Propagation}(b). The shape retention is a consequence of the low intensity part of the PDF ($I/\langle I \rangle\leq 1.5$) remaining almost flat, while the increased maxima irregularity is reflected in the loss of the sharp cutoff at high intensity values ($I/\langle I \rangle\geq 1.5$) of the PDF. While the spectrally detuned pattern does not strictly adhere to the customized PDF, it is still non-Rayleigh. With greater spectral detuning Fig.~\ref{Propagation}(c-e) the speckles gradually revert and become Rayleigh when the spectral decorrelation length $\Delta \lambda$ = 4 nm is reached (Fig.~\ref{Propagation}(e)). For further increases in the detuning the outputs adhere to Rayleigh statistics. 

An important conclusion to draw from Fig.~\ref{MonoPDF} is that monochromatic speckle customization influences the statistics of other wavelengths within the spectral coherence length. Specifically, the functional form of the intensity PDF and the speckle topology are largely preserved within a quarter of the spectral decorrelation length, $\Delta \lambda = \pm 1$ nm. This has two implications for broadband speckle customization, which are explored in the next two sections. First, if different intensity PDFs are desired at different wavelengths, in principle the PDFs can be arbitrarily chosen if their wavelengths are separated by more than the spectral decorrelation length. If the wavelength separation is less, however, the PDF choice will face constraints. For example, the intensity moments of the PDFs, $\langle I^{n}\rangle = \int_{0}^{\infty} P(\tilde{I}) \tilde{I}^{n} d\tilde{I}$, may need to have similar values depending on the exact wavelength separation. Second, when performing broadband speckle customization, targeting the same PDF across the pulse spectrum, the statistics of the speckles within a quarter of the spectral decorrelation length, $\Delta \lambda = \pm 1$ nm, do not need to be directly modified.

\subsection*{Multichromatic Customization}

\begin{figure}[hthb]
    \centering
    \includegraphics[width=3 in]{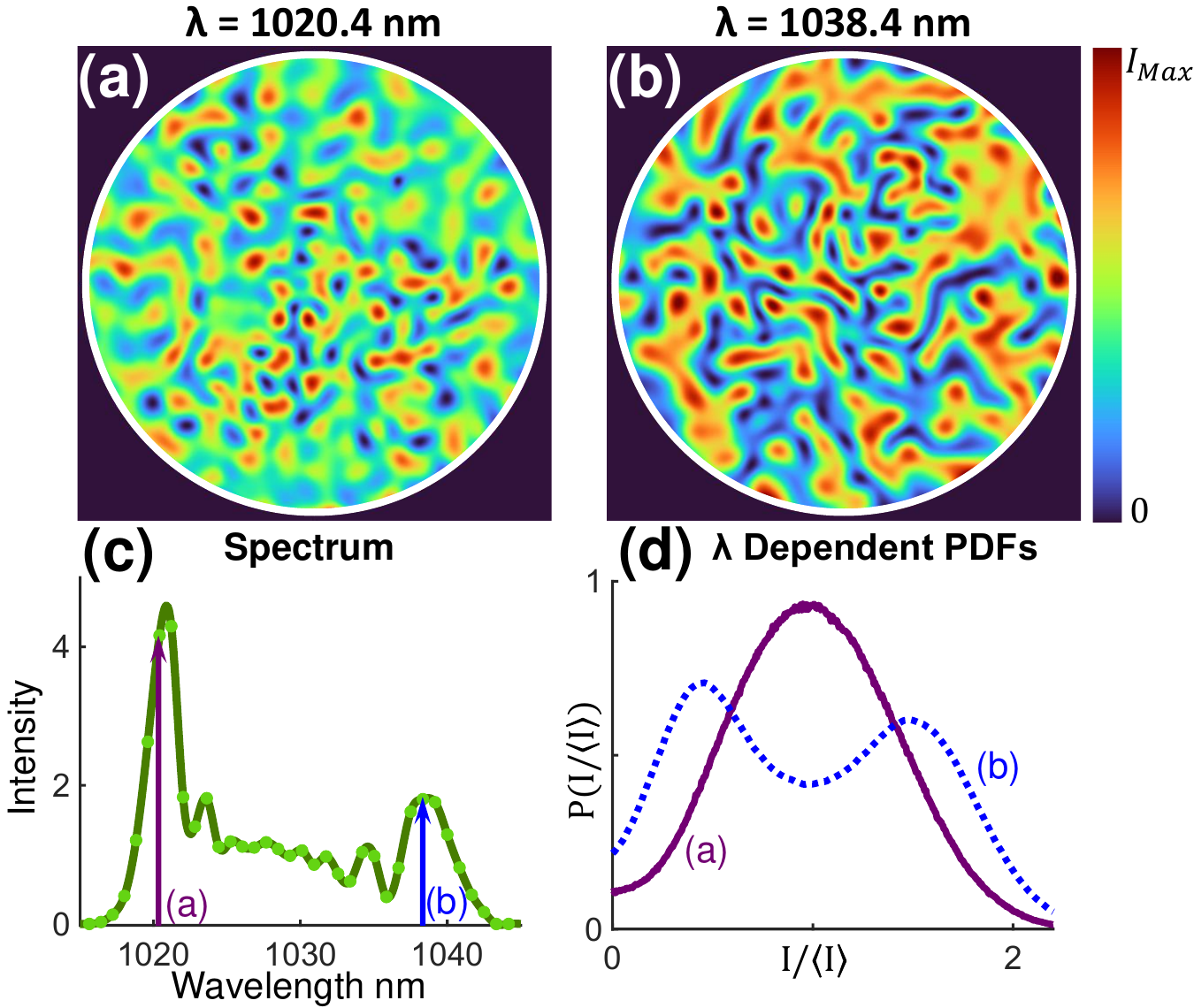}
    \caption{\textbf{Independent customization of  the intensity PDFs at different wavelengths, with the same SLM pattern.} Output intensity patterns at 1020.4 nm (a) and 1038.4 nm (b) -- opposite sides of the spectrum (c) -- are tailored to have unimodal and bimodal intensity PDFs (d) respectively.  The white circle is the boundary of the  fiber core. One hundred independent speckle patterns are used to calculate each PDF, and the experimental data are obtained from the measured field mapping matrix.}
    \label{PolyPDF}
\end{figure}

By incorporating a second wavelength into the field-mapping matrix used in the algorithm, we can arbitrarily customize the intensity statistics of speckle patterns at different wavelengths simultaneously. With a single SLM field modulation pattern, the field output from the fiber can have distinct intensity statistics at different wavelengths. As an example, a single speckled output designed to obey a unimodal intensity PDF at 1020.4 nm while simultaneously obeying a bimodal intensity PDF at 1038.4 nm is shown in Fig.~\ref{PolyPDF}. The spectral separation of these outputs, 18 nm, is approximately the pulse bandwidth (Fig.~\ref{PolyPDF})(c)). Between the two customization wavelengths, the speckles follow Rayleigh statistics in accordance with the detuning behavior shown in Fig.~\ref{Propagation}. In supplementary section 4 we show that as long as the wavelength separation is equivalent to or greater than the spectral decorrelation length, the wavelengths can be arbitrarily chosen with no effect on the intensity PDFs. Furthermore, we have not observed any restrictions on the choice of PDFs when tailoring speckles at two wavelengths, relative to one wavelength, under the same condition.

Comparison of the ensemble average intensity PDFs in Fig.~\ref{PolyPDF}(d) with Figs.~\ref{MonoPDF}(d,e) does not indicate any major changes in our ability to control the intensity PDF when increasing to multiple wavelengths. The unimodal intensity PDFs presented in the two figures are almost indistinguishable, a trend generally seen with other intensity PDFs generated. The bimodal intensity PDF in Fig.~\ref{PolyPDF}(d) does exhibit slight differences from the PDF in Fig.~\ref{MonoPDF}(e). Specifically, in Fig.~\ref{MonoPDF}(e) the maxima of the peaks in the PDF are approximately twice the value of the probability density at $I/\langle I \rangle =1$, while in Fig.~\ref{PolyPDF}(d) they are approximately 1.75 times as large as $P(I/\langle I \rangle =1)$. Furthermore, in the multichromatic customization case the bimodal intensity PDF has a more prominent high-intensity tail than the monochromatic case. Differences arise for the bimodal case, and not the others, because our algorithm can’t create a speckle pattern that adheres to the ``impossible” intensity PDF given to the algorithm, as discussed previously. Factors such as the ratio of peak intensities and the high-intensity tail are good qualitative measures for changes in our ability to customize the intensity statistics of a speckle pattern. These differences indicate that customizing the speckles at two wavelengths incurs only a small reduction in the degree of control that is possible.

\subsection*{Full-Spectrum Customization}

\begin{figure}[hthb]
    \centering
    \includegraphics[width=3 in]{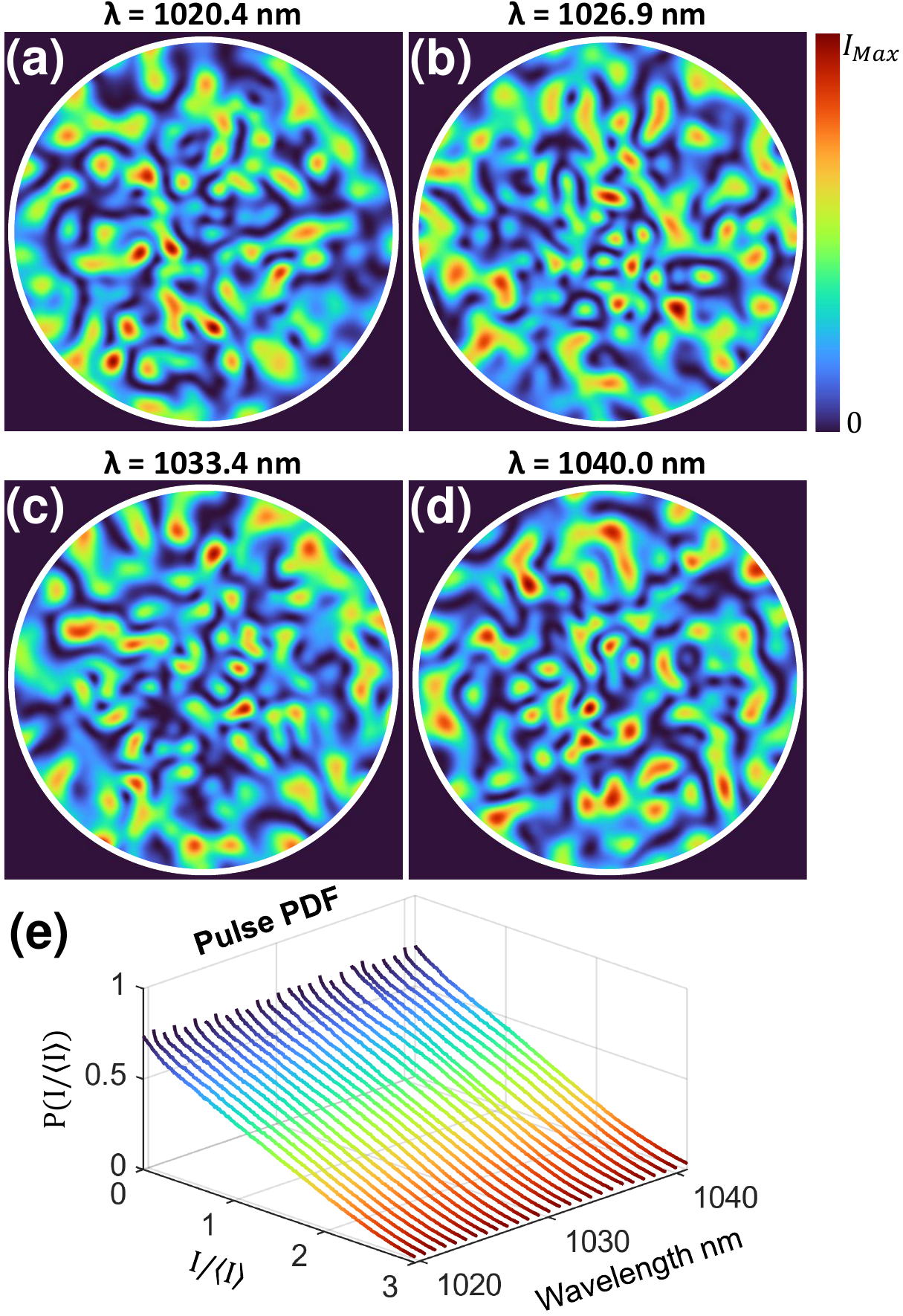}
    \caption{\textbf{Full spectrum PDF customization.} For a single SLM pattern, every wavelength of the pulse has an output (a-d) with an intensity PDF tailored to decrease linearly (e) rather than exponentially. The white circle is the boundary of the fiber core,  One hundred independent speckle patterns are used to calculate each PDF, and the experimental data are obtained from the measured field mapping matrix.}
    \label{PulsePDF}
\end{figure}

 The experimental limit of our ability to control spatio-spectral intensity statistics is illustrated in Fig.~\ref{PulsePDF}, where we tailor the speckles to obey non-Rayleigh statistics across the entire pulse spectrum (1019.6 nm to 1040.8 nm). Specifically, we tailor the speckles to have a linearly-decreasing intensity PDF at every wavelength in the pulse output from the fiber, like the PDF shown in Fig.~\ref{MonoPDF}(a). In Figs.~\ref{PulsePDF}(a-d) different custom monochromatic speckles output from the fiber are shown, generated by a single SLM pattern. The customized speckles have completely different spatial profiles, since they are separated by more than one spectral decorrelation length. However, the intensity PDFs (Fig.~\ref{PulsePDF}(e)) are the same across the entire spectrum. To accomplish this, we use every-other spectral component of the measured field-mapping tensor (approximately 1.6 nm separation per wavelength) in the optimization matrix. When plotting the PDFs in Fig.~\ref{PulsePDF}(e), we use a 0.8-nm spectral resolution. Because the PDFs in Fig.~\ref{PulsePDF}(e) are the same, we know that the intensity statistics apply to every wavelength within the range.

In contrast to the two-wavelength optimization (Fig.~\ref{PolyPDF}), in  full-spectrum optimization there are limits on what PDFs can be created. While we can generate unimodal intensity statistics across the spectrum (see supplementary section 5), we cannot create bimodal intensity statistics across the spectrum. To understand why this is the case, consider the relevant degrees of freedom. The SLM has 156 macro-pixels that can modulate the field incident on the fiber input. For a single polarization the disordered fiber has about $325$ spatial modes, and 5 spectrally distinct outputs (the pulse bandwidth divided by the spectral decorrelation length). Thus, the field we are attempting to control has approximately 1625 degrees of freedom, while the SLM has 156 degrees of control. Given the order-of-magnitude difference, it is not surprising that the full-spectrum customization cannot be done arbitrarily. In fact, it is remarkable that we can still manipulate properties of the PDF, such as the general shape (linear decay or unimodal). Furthermore, because we still possess some control over the shape of the PDF we retain the ability to control properties such as the speckle contrast, $C=\sqrt{\langle I^2 \rangle/\langle I \rangle^2 - 1}$, smililarly to what was shown in~\cite{bromberg2014generating} for monochromatic light, which is a sufficient degree of control for many exciting applications of tailored speckles.

\section*{Conclusion}

In this work, important questions related to the generation of customized random light have been answered. First, we demonstrated that it is possible to arbitrarily tailor the statistics of monochromatic random light at the output of a complex medium (a disordered fiber). Not only was this done with fewer degrees of control over the input (156 SLM macro-pixels) than  degrees of freedom in the output field (325 modes for one wavelength and polarization), but this arbitrary control was exerted over the entire output of the fiber core. Previous techniques for monochromatic speckle customization were either restrictive in terms of the obtainable statistics \cite{bromberg2014generating, dogariu2015electromagnetic, fischer2015light, amaral2015tailoring, kondakci2016sub, di2016tailoring, guillon2017vortex,  di2018hyperuniformity, devaud2021speckle, liu2021generation}, or could generate arbitrary statistics in free space but only over a target region consisting of about 1/4 of the total far-field controlled by the SLM~\cite{bender2018customizing, bender2019introducing, bender2019creating, bender2021circumventing, han2023tailoring}. Second, we demonstrated that with a single SLM pattern, two spectrally-decorrelated outputs from the fiber can be generated with different arbitrary intensity statistics. Given that the generally-accepted mechanism behind the creation of tailored speckle patterns is the presence of higher-order (4th, 6th, 8th, etc…) correlations in the random partial waves that interfere to create the pattern~\cite{bromberg2014generating, bender2018customizing, bender2019creating}, it was not obvious that this could even be done for the same statistics at decorrelated wavelengths -- much less different statistics-- since the wavelength-dependent field mappings are completely different. Finally, we have demonstrated that it is possible to manipulate the speckle statistics across the entire pulse spectrum at the output of the fiber simultaneously.

\begin{backmatter}
\bmsection{Funding} This work was funded by the Department of the Navy, Office of Naval Research (N00014-20-1-2789) and the National Science Foundation (ECCS-1912742). 

\bmsection{Acknowledgments} NB would like to thank SeungYun Han and Hui Cao for insightful discussions on speckle statistics.

\bmsection{Disclosures} The authors declare no conflicts of interest.

\bmsection{Data availability} Data underlying the results presented in this paper are not publicly available at this time but may be obtained from the authors upon reasonable request.

\end{backmatter}

\bibliography{References}

\newpage

\section*{Supplementary Information: Spectral Speckle Customization}

\section*{Field Mapping Operator}

Our method for generating customized speckles relies on the use of an experimentally measured field-mapping operator that can predict the 3D-field ($\hat{x}, \hat{y}, \hat{\lambda}$) output from the disordered fiber, when any field modulation pattern is displayed on the SLM. To obtain this operator we use the following approach.

\subsection*{Field Measurement}
To obtain the 3D field associated with a single SLM modulation pattern, we use off-axis digital holography~\cite{Cuche2000Spatial} with a known reference pulse. The reference pulse is taken directly from our laser source -- a train of approximately transform-limited pulses -- and illuminates the camera at an angle relative to the light output from the disordered fiber. The reference beam is expanded so that it can be approximated as a spatial plane-wave. An optical delay line allows matching of the signal and reference path lengths as well as changing the arrival time of the reference pulse. Our delay stage can vary the reference pathlength by a total of 25 cm ($\pm 12.5$ cm about the zero-point delay) which equates to 0.834 ns ($\pm 0.417$ ns about the zero-delay point) in the time domain. By scanning the delay of the reference pulse $\pm 2.16 $ ps about zero delay and imaging the 2D interference pattern measured by the camera every 2.66 fs, we can construct a 3D fringe pattern ($\hat{x}, \hat{y}, \hat{t}$) that encodes the interference between the reference pulse and the signal pulse. Using standard digital holography in the ($\hat{x}, \hat{y}$) plane, a temporal Fourier transform, and the convolution theorem to remove the reference, we can construct the 3D field ($\hat{x}, \hat{y}, \hat{\lambda}$) output from the fiber for a given SLM field modulation pattern.

\subsection*{Field-Mapping Operator Construction}
Using the field measurement described in the previous section, we can construct the desired field-mapping operator that characterizes the disordered fiber, using a process similar to a conventional transmission matrix measurement~\cite{popoff2010measuring}. By displaying an orthogonal set of basis vectors (field-modulation patterns) on the SLM and recording the 3D fields ($\hat{x}, \hat{y}, \hat{\lambda}$) output from the fiber, we can construct a field mapping operator that can predict the 3D field output from the fiber associated with any field modulation pattern displayed on the SLM. 

\begin{figure}[hthb]
    \centering
    \includegraphics[width=\linewidth]{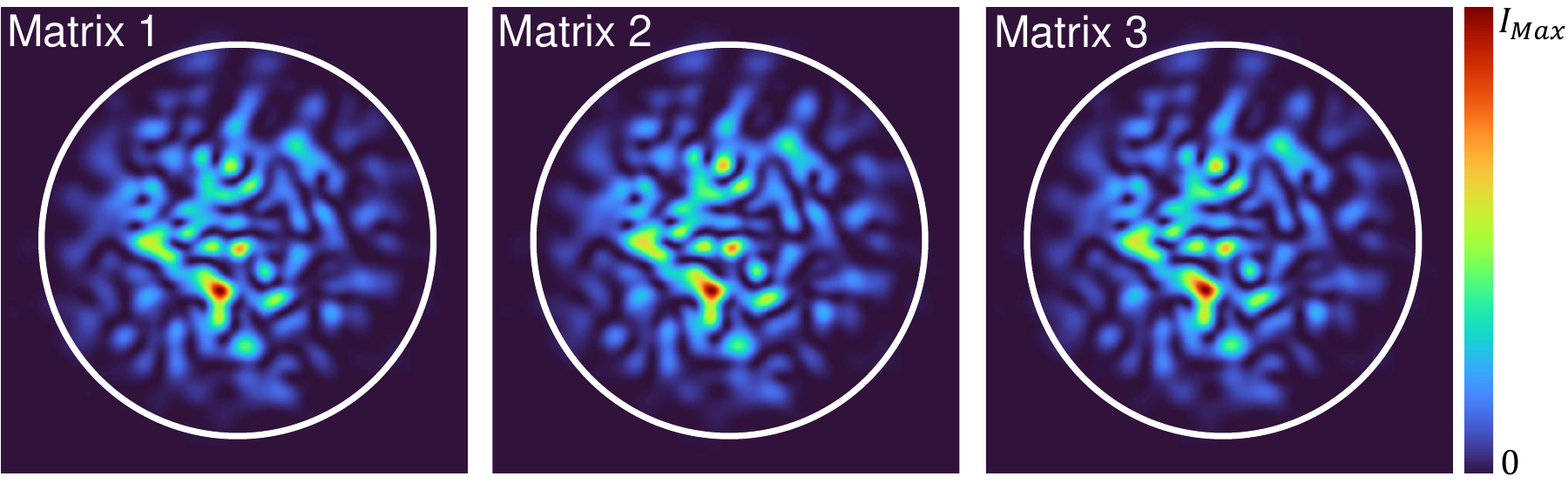}
    \caption{\textbf{Independent field-mapping measurement comparison.} Example speckled intensity outputs from three independently measured field mapping matrices are shown, demonstrating the agreement between the operators, $|C_{Ave}| = 0.97$.}
    \label{MatrixV}
\end{figure}

The choice of basis is important when performing this measurement. Naively, one may be inclined to use the individual macro-pixels on the SLM as a basis. This would be done by sequentially diffracting each macro-pixel to the first-order, while having the rest set to the zeroth-order. This set of basis vectors can be represented by the identity matrix. In practice, the identity matrix basis is disadvantageous because of its relatively low signal to noise. Typically, the Hadamard basis is used instead: defined by a square Hadamard matrix of dimension $2^k$, which can recursively be defined by ${\bf H}_{1}=1$,
\begin{equation}
{\bf H}_{2} = 
\begin{pmatrix}
1 & 1  \\
1& -1\\ 
\end{pmatrix},
\end{equation}
and
\begin{equation}
{\bf H}_{2^k} = 
\begin{pmatrix}
{\bf H}_{2^{k-1}} & {\bf H}_{2^{k-1}} \\
{\bf H}_{2^{k-1}} & -{\bf H}_{2^{k-1}}\\ 
\end{pmatrix}.
\end{equation}
Specifically in this work, we use a random Hadamard matrix to define our basis vectors. A random Hadamard matrix can be created by multiplying a regular Hadamard matrix by a diagonal matrix with random phases, $e^{i\theta}$, in each nonzero element (see~\cite{bender2021controlling} for more details). The advantage of this basis is that by changing the random phases we can create independent random Hadamard matrices. By measuring the system with multiple independent random-Hadamard matrices (i.e., independent sets of SLM patterns) we can detect both systematic and non-systematic error in the measurements. This is important because a single matrix measurement can take over one day. Furthermore, we can average over multiple measurements to further improve the signal to noise ratio of the final operator. The measured field-mapping operator used in this work was constructed using 3 independent random-Hadamard matrices. The average field correlation (Pearson correlation) of combinations of the 3 operators has magnitude $|C_{Ave}| = 0.97$, without any data processing or noise removal. Given that $|C_{Ave}| = 1$ indicates perfect agreement, we are satisfied that our measured operators can predict the output field associated with any field modulation pattern displayed on the SLM. In Fig.~\ref{MatrixV}, a representative output speckle pattern -- generated by each of the matrices using the same random SLM pattern -- is shown to highlight the agreement among the three operators. Given the level of agreement, we can rely on the measured operator to predict the field output from the fiber for any field-modulation pattern displayed on the SLM (see~\cite{bender2019creating} for an equivalent comparison). Because of this agreement, and the long measurement times of our system, the experimental data presented in this work are generated using the measured field mapping matrix. 

Once we obtain the averaged field-mapping matrix, we reject some of the SLM macro-pixels. The rationale behind this is as follows. When defining our basis, we use random-Hadamard matrices with $2^8$ elements in each column/row. The easiest way to map this basis to the SLM liquid crystal display is via a $16 \times 16$ macro-pixel grid, where the width of the grid approximately matches the pupil-diameter of the objective lens that couples light into the fiber. As a result, however, the pixels in the corners do not appreciably couple light into the fiber. Thus, we ignore the pixels outside a circle with a radius of $\approx 7$ macro-pixels, centered around the highest transmitting SLM macro-pixel, when constructing the final operator. This restriction leaves us with 156 macro-pixels, but they are responsible for $95\%$ of the light coupled into the fiber relative to the 256 pixels measured. This reduction is important because we calculate the pseudo-inverse operator in our algorithm (noisy singular values can corrupt the calculation of this operator). 

\newpage
\section*{Effect Of Normalization}
 
\begin{figure}[t]
    \centering
    \includegraphics[width=3 in]{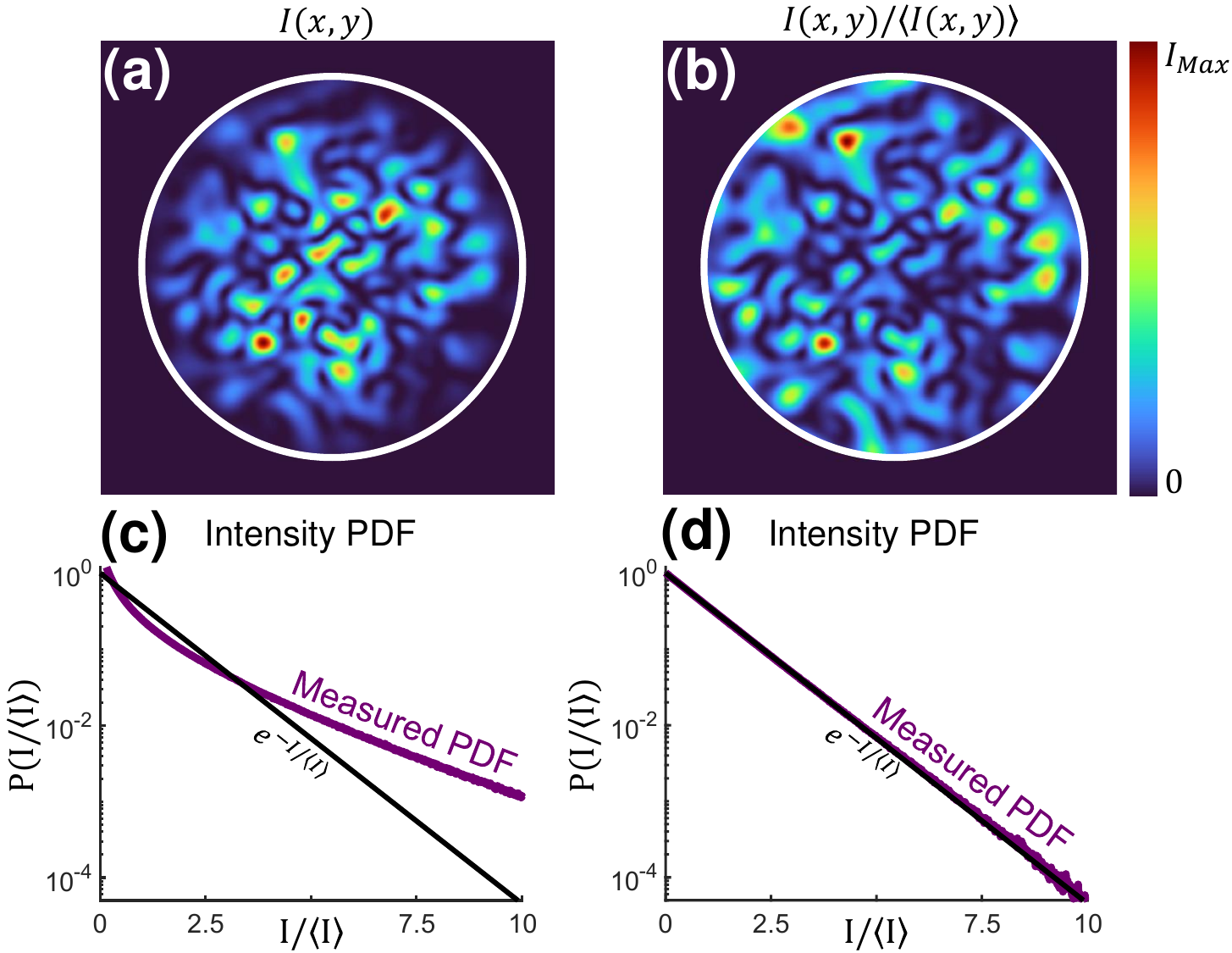}
    \caption{\textbf{Unnormalized vs. normalized speckles.} An example unnormalized speckle pattern (a) output from the fiber is juxtaposed with its normalized spatial profile (b). The differing intensity PDFs of the unnormalized (c) and normalized (d) ensembles of speckles are shown. The intensity PDFs are calculated using 1000 patterns (purple line) and compared with a Rayleigh PDF (black line) in (c,d). In (a,b) the white circle denotes the fiber core.}
    \label{UnNorm}
\end{figure}

In the main text, we always normalize the spatial intensity patterns using the convention $\overline{I(x,y)/\langle I(x,y) \rangle}$ = 1, where $\langle~\cdots~\rangle $ denotes ensemble averaging and the overline indicates spatial averaging. Specifically, this normalization ensures that the speckle statistics at every point in space and in every randomly generated pattern are the same, namely Rayleigh. Because we are using graded index fiber, however, $\langle I(x,y) \rangle$ is not uniform across the fiber core. As a result, the statistical properties of a given unnormalized speckle pattern (i.e., when we use $\overline{I(x,y)}$ = 1 instead of $\overline{I(x,y)/\langle I(x,y) \rangle}$ = 1), are not necessarily the same as the statistics at a given point in space (Rayleigh) when calculated using ensembles of independent speckles. In Fig.~\ref{UnNorm} we illustrate this by juxtaposing unnormalized and normalized output speckle patterns. The ensemble-averaged intensity PDF of the unnormalized speckles, (purple line) in Fig.~\ref{UnNorm}(c)), deviates from Rayleigh statistics (black line). Conversely, the normalized speckles’ ensemble-averaged intensity PDF (Fig.~\ref{UnNorm}(d)) exhibits Rayleigh statistics.

\begin{figure}[h]
    \centering
    \includegraphics[width=\linewidth]{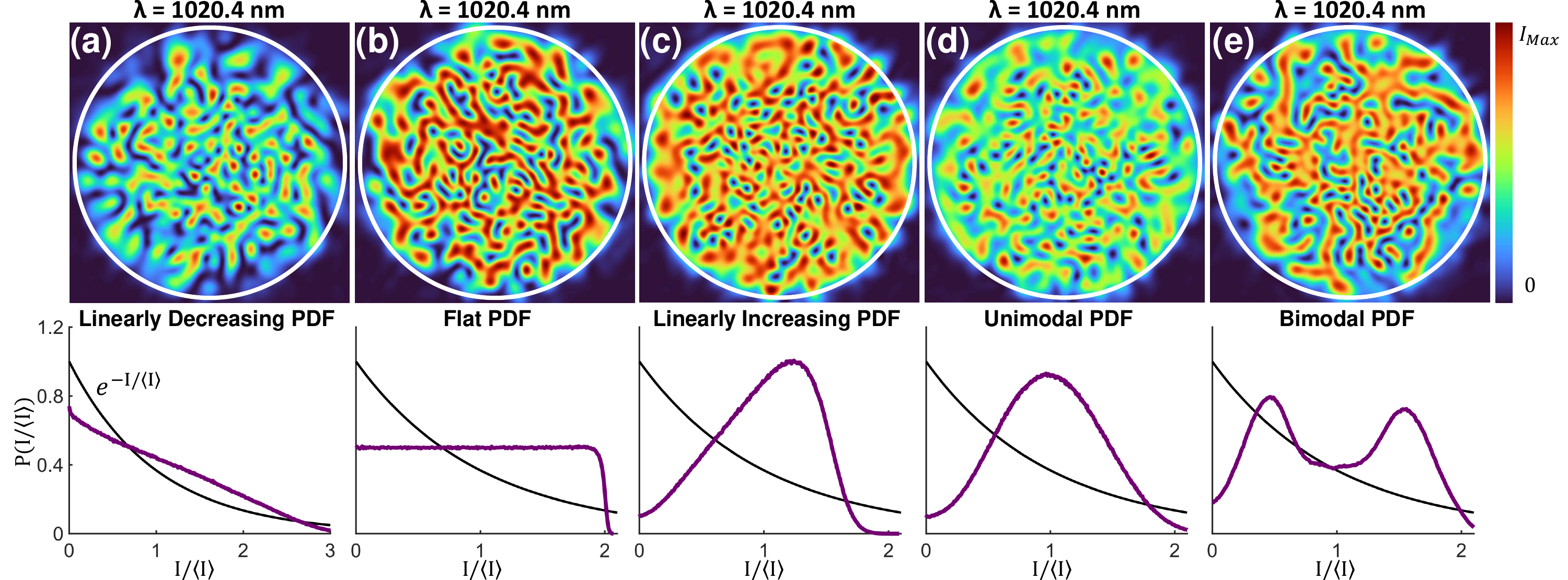}
    \caption{\textbf{Unnormalized monochromatic speckle patterns -with tailored intensity PDFs- equivalent to those shown in Fig. 2 of the main text.} Output speckle patterns adhering to linearly decreasing (a), uniform (b), linearly increasing (c), unimodal (d), and bimodal (e) intensity PDFs are shown. Below each output pattern, the associated ensemble averaged intensity PDF (purple line) is compared with a Rayleigh PDF (black line). Note, the white circled marks the fiber core, 100 independent speckle patterns are used to calculate each PDF, and the experimental data is generated using the measured field mapping matrix.}
    \label{UnNormCustom}
\end{figure}

It is conceivable that in some instances modification of the statistics of unnormalized speckles will be of interest. Since our speckle customization technique is independent of the initial PDF (it is adaptive), we can customize unnormalized speckles with the same facility as normalized speckles.  In Fig.~\ref{UnNormCustom}, we demonstrate the same ability to customize unnormalized speckles that was shown for normalized speckles in Fig.~2 of the main text. When generating the ensembles for Fig.~\ref{UnNormCustom}, we use the same algorithm parameters as those used to create Fig.~2: the desired PDF functional form and number of iterations. The only noticeable difference between the normalized/unnormalized cases is for  the case of a  linearly-increasing intensity PDF, presented in panel (c) in both Figs.~\ref{UnNormCustom} \&~2. For the unnormalized case, the drop-off after the maximal intensity value is more gradual than in the normalized case. This is the result of using the same number of iterations in an algorithm with two different initial conditions, and as such, is not significant. 

\newpage

Finally, it is worth mentioning that the speckles in Fig.~\ref{UnNormCustom} are not statistically stationary and ergodic, while those shown in the main text are. Specifically, the customized speckles still have an ensemble averaged mean intensity $\langle I(x,y) \rangle$ that varies as a function of space. This is because the speckles we initialize the algorithm with have this property, and we use a local intensity transformation which cannot necessarily remove it.

\newpage
\section*{Field Probability Density Functions}

\begin{figure}[h!]
    \centering
    \includegraphics[width=3 in]{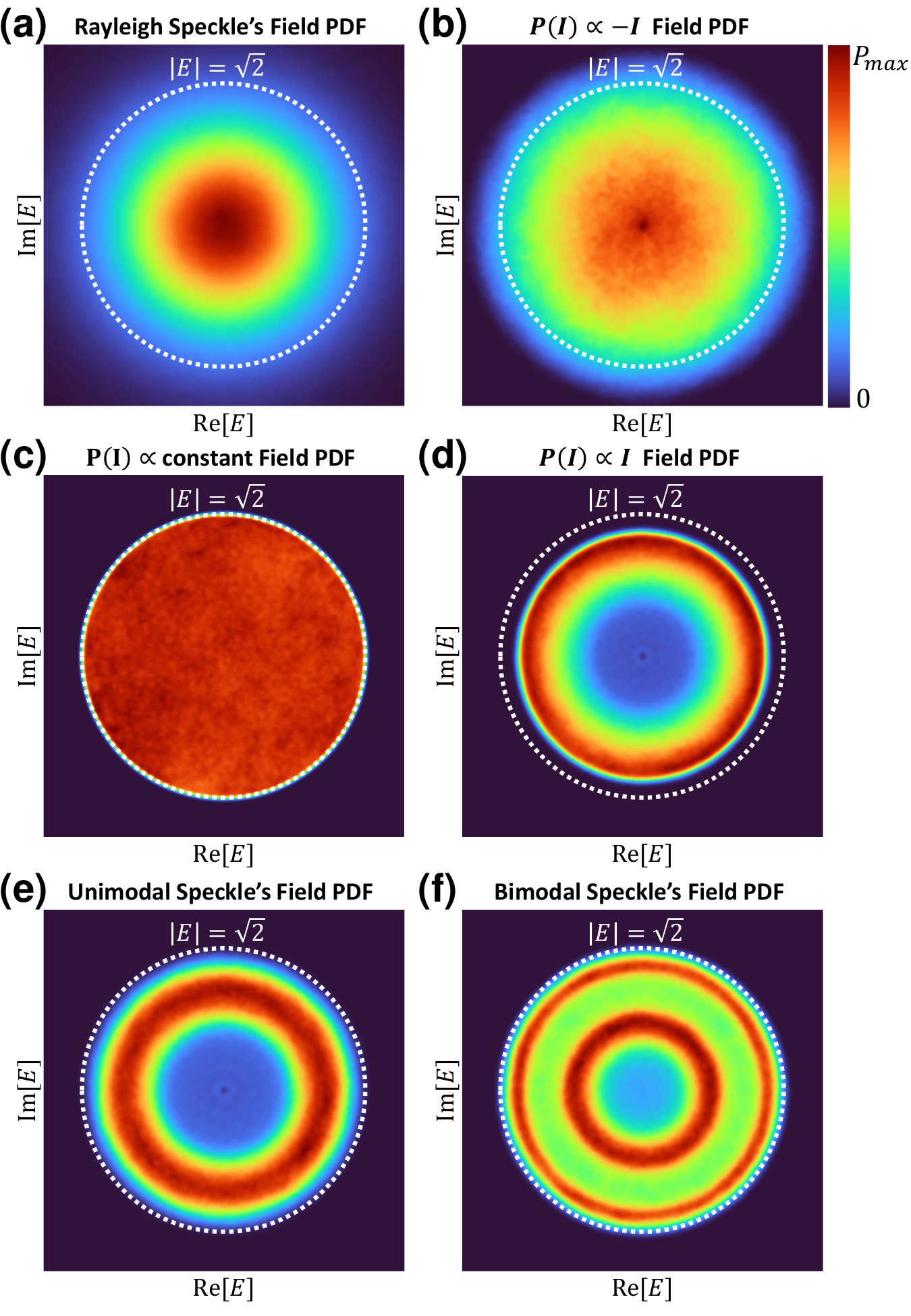}
    \caption{\textbf{Field probability density functions.} The joint-PDFs of speckled fields adhering to the following intensity PDFs are shown: Rayleigh (a), linearly decreasing (b), uniform (c), linearly increasing (d), unimodal (e), and bimodal (f). An ensemble of 1000 speckle patterns is used in (a), and ensembles of 100 are used in (b-f).}
    \label{FieldPDF}
\end{figure}

One of the important requirements when generating customized speckles is that they are fully-developed. Qualitatively, a fully-developed speckle pattern can be thought of as having a fully-randomized field. In most speckle pattern applications, this is either desired or necessary. For a field to be fully-developed, the amplitude and the phase values of the field must be statistically independent (i.e. the phase and amplitude values at a given point are independent of each other and their neighboring values beyond the speckle coherence length), and for the phase PDF to be uniformly distributed between 0 and $2\pi$. To verify that the speckles generated by our method are fully-developed, in Fig.~\ref{FieldPDF} we plot the field PDFs of the ensembles of speckle patterns shown in Fig.~2 of the main text. In each PDF, the functional form along the radial axis is the same for every angle. This property, known as circularity, indicates that the speckles are fully-developed. 

\newpage
\section*{Multichromatic Customization}

\begin{figure}[h!]
    \centering
    \includegraphics[width=3 in]{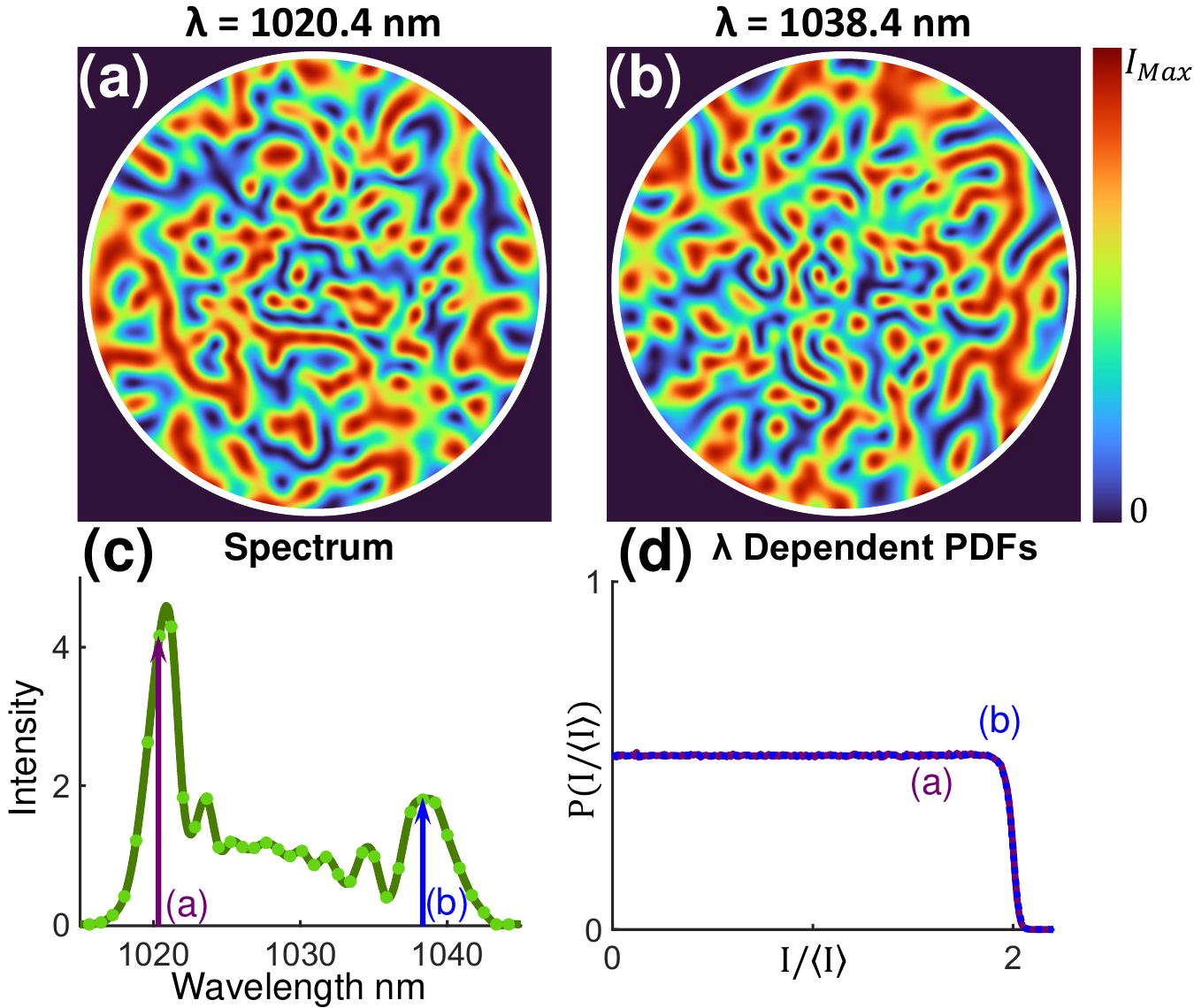}
    \caption{\textbf{Customizing  the intensity PDFs at different wavelengths, with the same SLM pattern.} Output intensity patterns at 1020.4 nm (a) and 1038.4 nm (b) -- opposite sides of the spectrum (c) -- are tailored to have flat intensity PDFs (d). Note, the white circled marks the fiber core, 100 independent speckle patterns are used to calculate each PDF, the experimental data is generated using the measured field mapping matrix, and in between the two wavelengths the speckles revert back to Rayleigh statistics.}
    \label{SamePDF}
\end{figure}

In the main text we demonstrated that different spectral components of a pulse (1020.4 nm and 1038.4 nm) can have different intensity PDFs for a single SLM field-modulation pattern (Fig.~4). In Fig.~\ref{SamePDF}, we demonstrate that this can be done for the same intensity PDF. At both wavelengths we design the intensity PDF to be constant. The nearly identical functional form of the intensity PDFs in Fig.~\ref{SamePDF}(d), when compared with Fig.~4(b), further supports the arguments made in the main text regarding the relatively low increase in difficulty when customizing the distributions at two wavelengths rather than one.

\begin{figure}[h!]
    \centering
    \includegraphics[width=3 in]{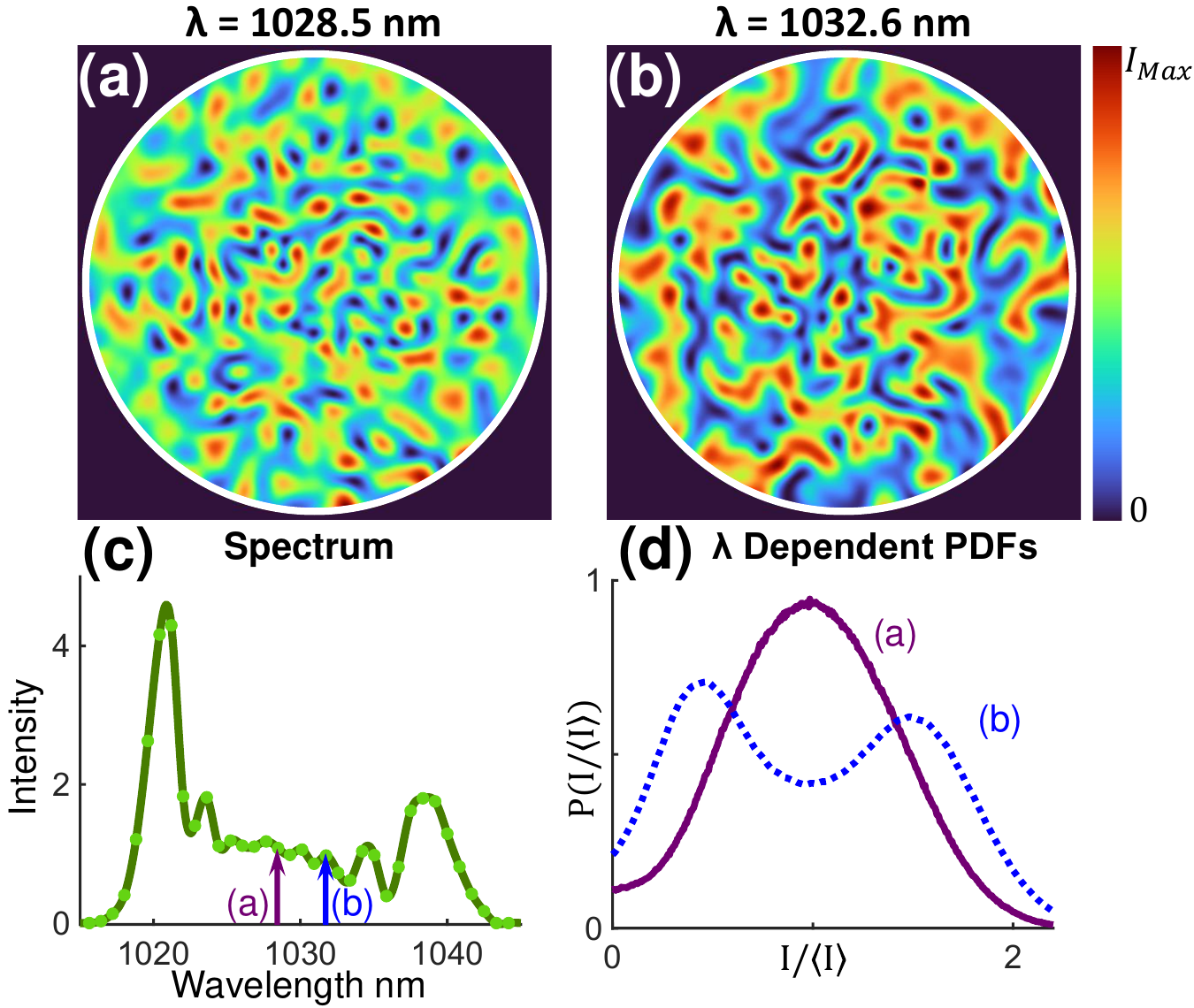}
    \caption{\textbf{Independently customizing  the intensity PDFs at different wavelengths, with the same SLM pattern.} Output intensity patterns at 1028.5 nm (a) 1032.6 nm (b) -- relative spectral position shown in (c) -- are tailored to have unimodal and bimodal intensity PDFs (d) respectively. Note, the white circled marks the fiber core, 100 independent speckle patterns are used to calculate each PDF, and the experimental data is generated using the measured field mapping matrix.}
    \label{PolyPDF_NP}
\end{figure}

We claim in the main text (Fig.~4) that the ability to tailor the intensity PDF is independent of the wavelengths chosen and their separation. The one caveat to this claim is that the wavelength separation must equal or exceed the spectral decorrelation length. In Fig.~\ref{PolyPDF_NP}, we customize speckle patterns at 1028.5 nm (a) 1032.6 nm (b) to have unimodal and bimodal statistics. Comparison between the PDFs in Fig.~\ref{PolyPDF_NP} (d), and the main text Fig.~4(d) reinforces this claim.

\newpage

\section*{Full-Spectrum Customization}

\begin{figure}[hthb]
    \centering
    \includegraphics[width=3 in]{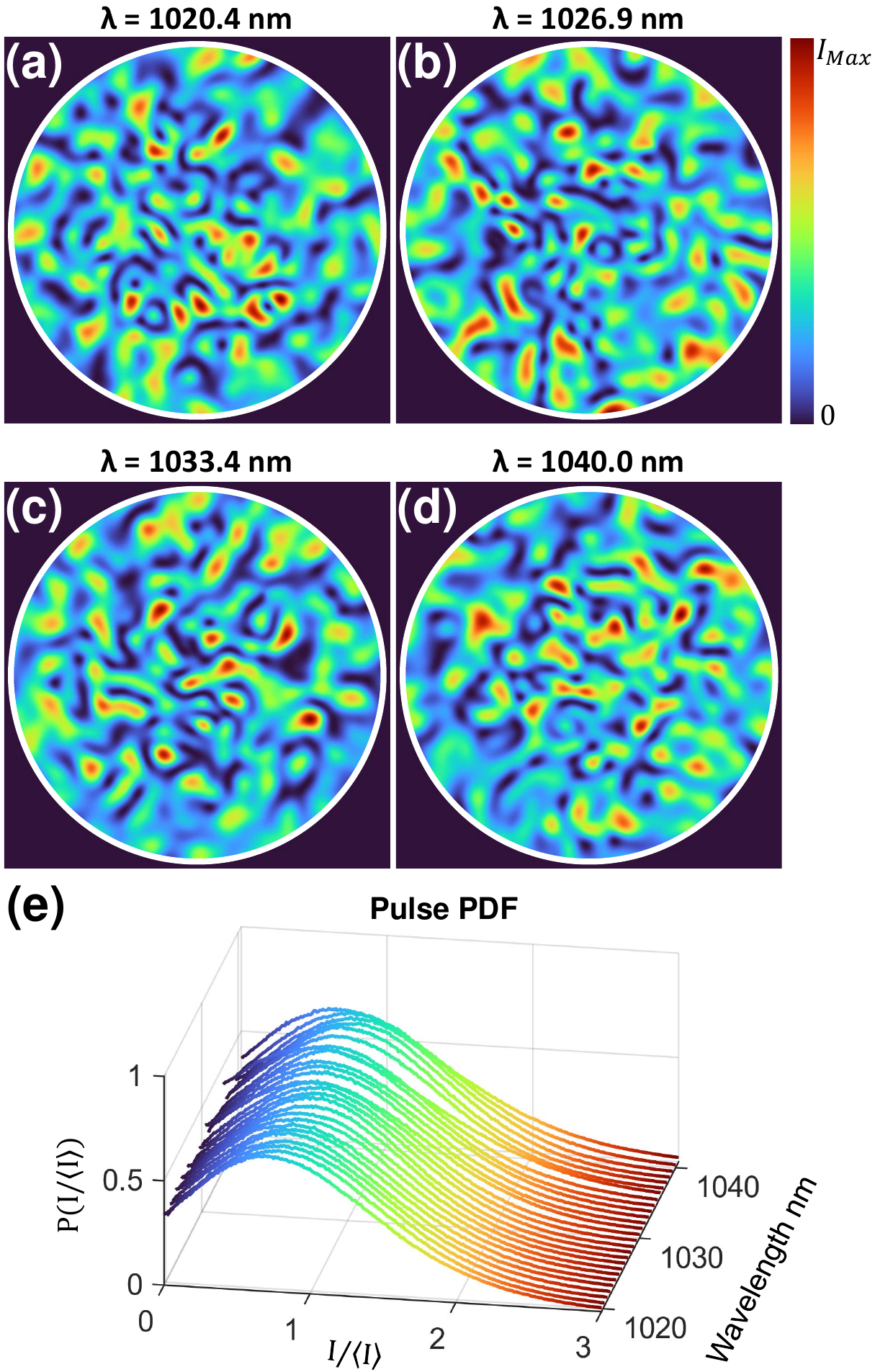}
    \caption{\textbf{Full pulse PDF customization.} For a single SLM pattern displayed, every wavelength of the pulse has an output (a-d) with a unimodal intensity PDF (e), instead of an exponentially decaying PDF. Note, the white circled marks the fiber core, 100 independent speckle patterns are used to calculate each PDF, and the experimental data is generated using the measured field mapping matrix.}
    \label{UniPulse}
\end{figure}

In the main text (Fig.~5) we demonstrate that all of the spatial-spectral components of the pulse can obey non-Rayleigh statistics (1019.6 nm to 1040.8 nm). Specifically, we tailored the speckles to have a linearly decreasing intensity PDF at every wavelength in the pulse output from the fiber. Here we show that this can also be done for a unimodal intensity PDF. In Figs.~\ref{UniPulse}(a-d) different custom monochromatic speckles output from the fiber, generated by a single SLM pattern, are shown. The customized speckles have completely different spatial profiles, since they are separated by more than one spectral decorrelation length. The intensity PDFs across the pulse shown in Fig.~\ref{UniPulse}(e), however, are the same. While the exact functional form of the PDFs in (e) does not match Fig.~2(d), the general characteristics of a unimodal PDF are still present. Unlike the example shown in the main text, this example demonstrates that when customizing the entire pulse some limitations exist. However, these can potentially be overcome by increasing the effective number of input degrees of freedom (using more macro-pixels on the SLM, multiple polarizations, optical tapering devices, etc.).  

\end{document}